\documentclass[aps,prl,twocolumn,superscriptaddress,10pt,longbibliography,footnoteinbib,floatfix]{revtex4-2}
\usepackage{ORI_Group_style}
\usepackage[english]{babel}
\usepackage{lipsum}

\newcommand{\dipoz}{m_z}

\hypersetup{colorlinks,linkcolor=red,citecolor=blue,urlcolor=blue}

\begin{document}

\title{Quantum Size Effects in the Magnetic Susceptibility of a Metallic Nanoparticle}

\author{M.~Roda-Llordes}
\email{marc.roda-llordes@uibk.ac.at}
\affiliation{Institute for Quantum Optics and Quantum Information of the Austrian Academy of Sciences, A-6020 Innsbruck, Austria}
\affiliation{Institute for Theoretical Physics, University of Innsbruck, A-6020 Innsbruck, Austria}

\author{C.~Gonzalez-Ballestero}
\affiliation{Institute for Quantum Optics and Quantum Information of the Austrian Academy of Sciences, A-6020 Innsbruck, Austria}
\affiliation{Institute for Theoretical Physics, University of Innsbruck, A-6020 Innsbruck, Austria}

\author{A.~E.~Rubio~López}
\affiliation{Institute for Quantum Optics and Quantum Information of the Austrian Academy of Sciences, A-6020 Innsbruck, Austria}
\affiliation{Institute for Theoretical Physics, University of Innsbruck, A-6020 Innsbruck, Austria}

\author{M.~J.~Martínez-Pérez}
\affiliation{Instituto de Ciencia de Materiales de Aragón, CSIC-Universidad de Zaragoza,  E-50009, Zaragoza, Spain}
\affiliation{Fundación ARAID, Avda. de Ranillas, E-50018, Zaragoza, Spain}

\author{F.~Luis}
\affiliation{Instituto de Ciencia de Materiales de Aragón, CSIC-Universidad de Zaragoza,  E-50009, Zaragoza, Spain}

\author{O.~Romero-Isart}
\email{oriol.romero-isart@uibk.ac.at}
\affiliation{Institute for Quantum Optics and Quantum Information of the Austrian Academy of Sciences, A-6020 Innsbruck, Austria}
\affiliation{Institute for Theoretical Physics, University of Innsbruck, A-6020 Innsbruck, Austria}
\date{\today}

\begin{abstract}
We theoretically study quantum size effects in the magnetic response of a spherical metallic nanoparticle (e.g. gold). 
Using the Jellium model in spherical coordinates, we compute the induced magnetic moment and the magnetic susceptibility for a nanoparticle in the presence of a static external magnetic field. Below a critical magnetic field the magnetic response is diamagnetic, whereas above such field the magnetization is characterized by sharp, step-like increases of several tenths of Bohr magnetons, associated with the Zeeman crossing of energy levels above and below the Fermi sea. We quantify the robustness of these regimes against thermal excitations and finite linewidth of the electronic levels. Finally, we propose two methods for experimental detection of the quantum size effects based on the coupling to superconducting quantum interference devices. 
 \end{abstract}

\maketitle

Metal clusters lie in between the realms of bulk metals and atoms~\cite{Halperin1986,deJongh1994}. 
Determining the size at which the metallic behaviour arises is not easy as it depends on which physical quantity one uses to define 'metallicity'.  Some properties, such as the crystal structure or average bonding distances, converge to bulk values surprisingly quickly
and, in the case of e.g. the electron density, even for clusters of less than $200$ atoms~\cite{Mulder1994}. 
By contrast, clear deviations from bulk values are found in other properties for much larger nanoparticles. 
Two notable examples are the large charging energy of nanoscopic metal clusters~\cite{kubo_electronic_1962} and the size-dependent optical absorption in plasmonic nanoparticles~\cite{Liz-Marzan2006}.
Many of these phenomena stem from the nanoscale spatial confinement of the electronic wave-functions. The resulting, atom-like discrete energy level structure gives way, as the particle size increases, to a bulk-like quasi-continuum band structure. In the intermediate regime, the response of metallic structures can still show evidences of finite energy gaps or correlations between different levels. Such \emph{quantum size effects} have a strong impact in, among others, the electric conductance through metallic nano-bridges~\cite{Krans1995}, or the temperature and magnetic field dependence of the specific heat in e.g. Pd clusters~\cite{volokitin_quantum-size_1996}.

In this Letter we study quantum size effects in the magnetic response of a spherical metallic nanoparticle and discuss their experimental observation. 
Motivated by previous theoretical approaches based on the simplest picture of spinless free electrons~\cite{Bivin1977,van_ruitenbeek_model_1991,van_ruitenbeek_size_1993}, which have been applied to explore the orbital magnetism~\cite{Leeuwen1992,van_ruitenbeek_model_1991} of metallic clusters, we describe the nanoparticle using the Jellium model. 
This model is simple enough to allow for a numerically exact solution, while complex enough to successfully describe quantum size effects such as e.g. the minima in ionization potential of Na and K clusters with full electronic shells~\cite{Ekart1984,deHeer1993}. The Jellium model
is a particularly good description for high-density materials with free-electron-like conduction bands dominated by $s$ orbital electrons, especially alkali metals~\cite{fetter_quantum_2003}. Since the high reactivity of these metals makes nanoclusters only available in jets~\cite{deHeer1993}, we focus on Au nanoparticles whose equilibrium magnetic response is easier to observe experimentally. 
Gold combines a simple electronic structure, dominated by the outer $6s^{1}$ orbitals~\cite{rangel_band_2012} with excellent chemical stability.
The Jellium model provides a good description for gold whereby relativistic effects ({\em e.g.} spin-orbit coupling) can be safely neglected~\cite{gomez_viloria_orbital_2018,mathur_persistent-current_1991}.
Furthermore, Au nanoparticles with well controlled and homogeneous sizes and shapes can be chemically synthesized, stored in the form of colloids for long periods of time~\cite{Liz-Marzan2006,Faraday1857,Scarabelli2014}, and individually transferred to an on-chip location (e.g. above a magnetic sensing device to measure their magnetic response) with the tip of an atomic force microscope~\cite{Wang2008}. 

We consider a rigid, spatially fixed metallic nanosphere of radius $R$ in the presence of a homogeneous magnetic field $\BB(\rr)=B_0\eb_z$.
We describe its internal electronic degrees of freedom using the Jellium model~\cite{fetter_quantum_2003,giuliani_quantum_2005,brack_physics_1993}, i.e. we describe the particle as an ensemble of $N$ free electrons under the influence of a positively charged background, whose effect is to create an infinite spherical potential well of radius $R$ for the electrons.
The only material-dependent free parameter within this model is the electronic density $N/V\equiv 3N/(4\pi R^3) $, usually given through the Wigner-Seitz radius $r_s \equiv R/N^{1/3}$~\cite{fetter_quantum_2003}.
The Hamiltonian of the nanoparticle within this model is given by $\Hop = \Hop_0 + \Hop_B$. 
The first term describes the dynamics of its electrons for $B_0=0$, namely $\Hop_0 = \sum_{i=1}^N \ppop_i^2/(2m_e) + U\Theta (|\rrop_i|-R)$ with $U\to \infty$.
Here $\Theta(x)$ is the Heaviside step function, $m_e$ the electron mass, and $\rrop_i= (\hat x_i, \hat y_i, \hat z_i)$ and $\ppop_i$ the position and momentum operator vectors of the $i$-th electron. 
As $\hat H_0$ is spherically symmetric we choose a single-particle eigenstate basis $\{\ket{n l m s}\}$ composed by eigenstates of the orbital and spin angular momentum operators, i.e., [$\hat{l}^2,\hat{l}_z,\hat{s}_z ]\ket{n l m s} = \hbar [ \hbar l(l+1),m,s]\ket{n l m s}$ with $n \in \mathds{N}$, $l \in \mathds{N}_0$, $m \in \mathds{Z}$ with $\abs{m} \leq l$, and $s= \pm 1/2$ respectively.
Importantly, the corresponding eigenenergies, given by $E_{n l} = E_0 u_{l n}^2$ with $E_0\equiv\hbar^2/(2 m_e R^2)$ and $u_{l n}$ the $n$-th zero of the spherical Bessel function of the first kind of order $l$ (see Supplemental Material~\cite{SM} for details), are highly discretized due to the electronic confinement within the nanoparticle \cite{kubo_electronic_1962,PERENBOOM1981173,Halperin1986,giuliani_quantum_2005}.
The term of the Hamiltonian describing the effect of the homogeneous magnetic field reads~\cite{SM,gomez_viloria_orbital_2018}
\begin{equation}
    \Hop_B = \hbar\omega_L \frac{\Lop_z + 2\Sop_z}{\hbar} + \frac{1}{2}m_e\omega_L^2\rhoop^2
    \label{eq:H}
\end{equation}
where $\omega_L \equiv e B_0/(2m_e)$, $e>0$ is the fundamental charge, $\Lop_z \equiv \sum_i \hat{l}_{z,i}$, $\Sop_z \equiv \sum_i \hat{s}_{z,i}$ and $\rhoop^2 \equiv \sum_i \pare{\hat{x}_i^2+\hat{y}_i^2}$.
We are interested in the regime of nanometer-size particles and moderate $B$-fields ($\eta \equiv e B_0R^2/\hbar \ll1$), which allows us to treat the second term of \eqnref{eq:H} perturbatively \cite{SM}.
We consider as unperturbed eigenstates the single particle eigenstates of $\Hop_0$, namely $\ket{n l m s}$, which are also eigenstates of the first term in \eqnref{eq:H}.
The eigenenergies up to first order in perturbation theory are hence given by
\begin{equation}
    \frac{E_{n l m s}}{E_0} = u_{l n}^2 + \eta (m+2s) + \frac{\eta^2}{4R^2} \bra{nlms}\rhoop^2\ket{nlms}.
    \label{eq:eigenenergies}
\end{equation}
The effect of the homogeneous magnetic field is thus to lift degeneracies in analogy with the Zeeman effect
and to add a small diamagnetic energy to all eigenstates.

We are interested in the induced magnetic dipole of the sphere, given by the operator~\cite{buhmann_macroscopic_2012} 
\begin{equation}
    \hat{\mm} = - \frac{\mu_B}{\hbar } \sum_{i=1}^N \pare{ \hat\rr_i\times \spare{\hat\pp_i + e\AB(\hat\rr_i)} + 2\hat \ss_i}
    \label{eq:magnetic_moment}
\end{equation}
where $\mu_B \equiv \hbar e/(2m_e)$ is the Bohr magneton and $\AB(\rr)$ the electromagnetic vector potential.
Owing to the symmetry of the system, the expected value of \eqnref{eq:magnetic_moment} for a thermal state of $\Hop$ at temperature $T$ is parallel to the homogeneous magnetic field, i.e., $\expect{\hat\mm } = \dipoz \eb_z $ with 
\begin{equation}
    \dipoz = -\mu_B \pare{ \frac{\expect{\Lop_z}}{\hbar} + 
    2\frac{\expect{\Sop_z}}{\hbar} + \frac{e B_0}{2\hbar} \expect{\rhoop^2}}.
    \label{eq:expectation_value_magnetic_moment}
\end{equation}
The first two terms contribute to the paramagnetic response of the system, as states with negative orbital and spin angular momenta are energetically favoured, see Eq.~\eqref{eq:eigenenergies}. 
Conversely, the third term in Eq.~\eqref{eq:expectation_value_magnetic_moment} corresponds to a diamagnetic contribution.

 \begin{figure}[t]
	\includegraphics[width=\linewidth]{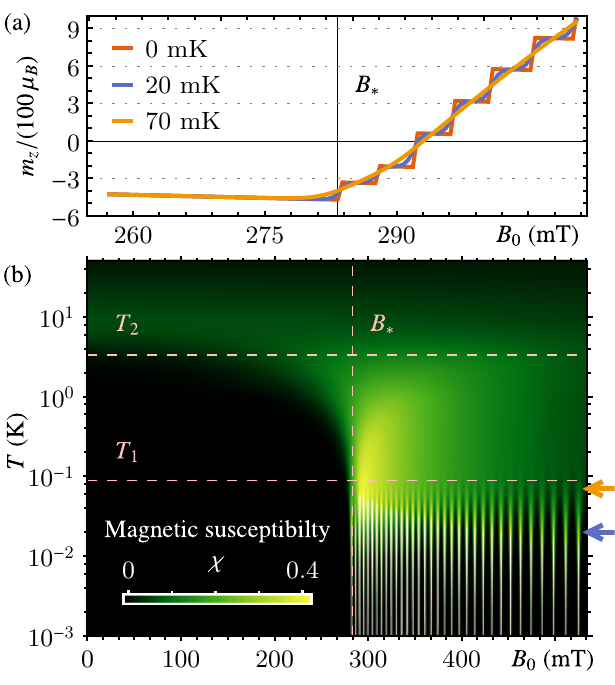}
	\caption{
		(a) Induced magnetic dipole moment $\dipoz$ [Eq.~\eqref{eq:expectation_value_magnetic_moment}] in a nanoparticle with radius $R=7.5$ nm as a function of applied magnetic field $B_0$, and at temperatures $T=0$ (red), $20$ mK (blue), and $70$ mK (yellow). 
		(b) Magnetic susceptibility $\chi$ [Eq.~\eqref{eq:chi_definition}] as a function of $B_0$ and $T$ for the same nanoparticle as above.
		The dashed lines indicate the critical field $B_*$ and temperatures $T_1$ and $T_2$, see main text for details.
		The colored arrows indicate the values of $T$ shown in panel (a).
		All panels correspond to gold ($r_s=3 a_0$).}
	\label{fig:magnetic_moment_results}
\end{figure}

Evaluating \eqnref{eq:expectation_value_magnetic_moment} requires calculating expected values of single-particle operators of the form~\cite{fetter_quantum_2003}
\begin{equation}
    \expect{\hat O} = \sum_{n l m s} \bar{n}_{n l m s}(T) \bra{nlms} o(\rrop,\ppop,\hat{\ss}) \ket{n l m s}.
    \label{eq:compute_observables}
\end{equation}
Here  $\bar{n}_{n l m s}(T)\equiv f_{\rm FD}(E_{nlms},\mu,T)$ is the mean thermal occupation of the electronic level $\ket{nlms}$, with $f_{\rm FD}(E,\mu,T) = \pare{1+\exp\spare{\pare{E-\mu}/\pares{k_B T}}}^{-1}$ the Fermi-Dirac distribution and $k_B$ the Boltzmann constant.
The chemical potential $\mu$ is defined implicitly via
\begin{equation}
    N = \pare{\frac{R}{r_s}}^3= \sum_{n l m s} \bar{n}_{n l m s}(T).
    \label{eq:fermi_energy_relation}
\end{equation}
At $T=0$ it defines the Fermi energy $E_F \equiv \mu(T=0)$, whereby $\bar{n}_{nlms}(T=0)= \Theta(E_F-E_{nlms})$.
To accurately model the physical conditions in the experimental proposal below, the number of electrons $N$ is assumed fixed. Thus,
to compute observables from Eq.~\eqref{eq:compute_observables}, one first needs to solve \eqnref{eq:fermi_energy_relation} to obtain the  $(N,T,B_0)$-dependent chemical potential $\mu(T,N,B_0)$.
Even in the simplest case $T=B_0=0$, this is an involved task as it requires to determine, for a given value of $N$ (i.e., for a given radius $R$), the lowest $N$ single-particle eigenenergies $E_{nlms}$.
This is difficult in practice for two reasons.
First, the small size of the particle discards a continuum limit approach.
Second, the spherical boundary conditions result in eigenenergies $E_{nlms}$ which, as opposed to other cases (\eg~cubic box boundary conditions~\cite{Bivin1977,van_ruitenbeek_model_1991,van_ruitenbeek_size_1993}) do not obey an absolute ordering rule, as the zeros of the spherical Bessel functions are interlaced in a complex way.
In other words, no absolute rule exists to determine which of two arbitrary zeros, $u_{ln}$ and $u_{l'n'}$, is the largest. 
As shown in detail in~\cite{SM}, we have developed a numerical algorithm that allows us to solve \eqnref{eq:fermi_energy_relation} exactly at $T=0$, and numerically with arbitrary precision at $T\ne0$, for the range of nanoparticle sizes we are interested in ($R \lesssim 40 \text{nm}$).

The induced dipole moment $\dipoz$, Eq. \eqref{eq:expectation_value_magnetic_moment}, is shown in \figref{fig:magnetic_moment_results}(a) as a function of external field $B_0$ for a gold nanoparticle ($r_s = 3a_0$ with $a_0$ the Bohr radius) with radius $R=7.5$ nm at three different temperatures. 
The induced moment shows two distinct behaviors delimited by a critical field $B_0=B_*$, to be defined and analyzed below. 
At fields $B_0<B_*$ the response is diamagnetic, $\dipoz \propto -B_0$, whereas at large enough fields $B_0 > B_*$ the magnetic moment not only starts to increase with $B_0$, but does so in quantized jumps of several tens of Bohr magnetons, equivalent in magnitude to sudden flips of several tenths of electronic spins at once.
This quantized behavior is more pronounced at zero temperature (red line in \figref{fig:magnetic_moment_results}(a)), and gradually smears away upon increasing $T$.
Note that the transition from decreasing to increasing behavior in $m_z$, at $B_0=B_*$, remains for higher temperatures.

The magnetic response of the nanosphere shown by \figref{fig:magnetic_moment_results}(a) can be understood from the single-particle eigenenergies in Eq.~\eqref{eq:eigenenergies}.
At $T=B_0=0$, the electronic ground state corresponds to the states with the lowest $N$ eigenenergies being occupied, and fulfils $\expect{\hat{L}_z} = \expect{\hat{S}_z} =0$ \footnote{This holds for values of $N$ resulting in the so-called full-shell electronic configuration~\cite{SM}. The generalization to other values of $N$ is straightforward and leads to a non-zero magnetization at zero field.}.
When a magnetic field $B_0$ is applied, the zero-field eigenenergies become non-degenerate, as states with different values of $m$ and $s$ experience a Zeeman splitting (see Eq.~\eqref{eq:eigenenergies}).
At weak fields $B_0<B_*$ these energy shifts are small, and states above and below the original Fermi level remain well separated.
The ground state is thus the same as at zero field, resulting in a diamagnetic response 
$\dipoz=- B_0 \mu_B e \expect{\hat{\rho}^2}/(2\hbar)$,
see Eq.~\eqref{eq:expectation_value_magnetic_moment}.
Conversely, at strong enough fields
$B_0>B_*$, the Zeeman-split states originally above and below the Fermi level start to overlap, and the ground state changes abruptly when a previously occupied state $\ket{n_1,l_1,m_1,s_1}$ becomes empty while a previously empty state $\ket{n_2,l_2,m_2,s_2}$ becomes occupied.
As a result, the induced magnetic dipole moment sharply changes by an amount $\mu_B\spare{(m_2+2s_2)-(m_1+2s_1)}$, typically of many Bohr magnetons due to the high values of the orbital quantum numbers $l$ and $m$ near the Fermi level.
Since $|s|\leq1/2$, the effects of the spin degree of freedom are negligible in comparison, as expected~\cite{von_oppen_magnetic_1994}.
At higher temperatures, the field-induced modification of the ground state is still present but the sharp changes in $\dipoz$ become less appreciable as the mean occupation number of states above and below the Fermi level is no longer $0$ or $1$ but given by the smoother function $\bar{n}_{n l m s}(T)$.
We remark that both the sudden slope change at $B_0=B_*$ and the step-like behaviour of $\dipoz$ are quantum size effects as they stem from the discrete character of the single-electron energy levels $E_{nl}\propto R^{-2}$ and the Pauli exclusion principle.

\begin{figure}[t]
	\includegraphics[width=\linewidth]{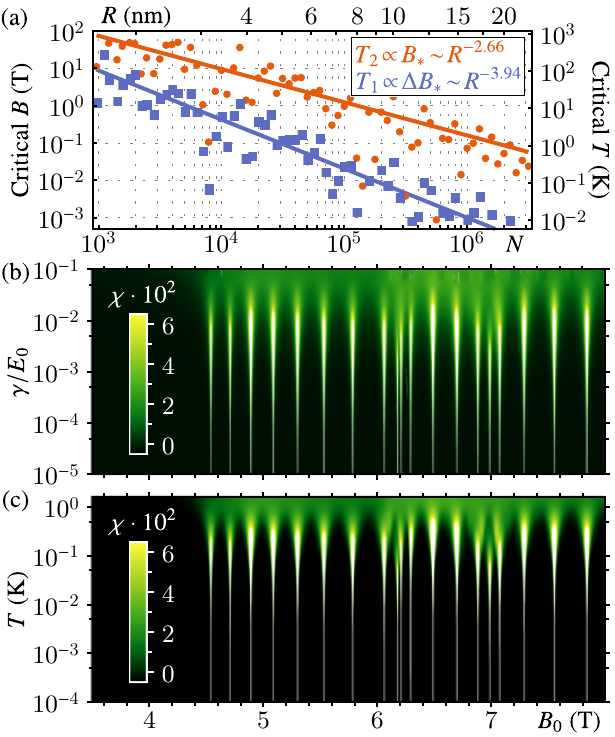}
	\caption{
		(a) Critical fields and temperatures $(B_*$, $T_2)$ (red) and $(\Delta B_*$, $T_1)$ (blue) as a function of electron number $N$ (lower axis) or particle radius $R$ (upper axis).
		Solid lines show the fit to a power law.
		(b) Zero-temperature magnetic susceptibility [Eq.~\eqref{eq:chi_definition}] of a nanoparticle with $R=3.7$nm
		as a function of applied field $B_0$ and linewidth of the single-electron levels $\gamma$.
		(c) Magnetic susceptibility as a function of temperature $T$ for $\gamma =0$ and the same nanoparticle as in (b).
		All panels correspond to gold ($r_s=3 a_0$).}
	\label{fig:BStar_and_Linewidth}
\end{figure}

In order to further characterize the magnetic response of the nanosphere, we compute the point-particle magnetic susceptibility~\cite{gomez_viloria_orbital_2018}, which we define as 
\begin{equation}
    \chi \equiv \frac{\mu_0}{V} \frac{\partial \dipoz}{\partial B_0}, 
    \label{eq:chi_definition}
\end{equation}
with $\mu_0$ the vacuum permeability.
We display $\chi$ in \figref{fig:magnetic_moment_results}(b) as a function of temperature $T$ and applied field $B_0$.
The magnetic response shows different regimes as a function of $T$ and $B_0$, delimited by three critical parameters, namely $B_*$, $T_1$, and $T_2$.
The critical field $B_*$, already introduced above, can now be rigorously defined as the field at which the first energy level crossing occurs at $T=0$. 
This corresponds to the first sharp peak in the susceptibility $\chi$ in the low temperature regime $T \rightarrow 0$.
Furthermore, we define two critical temperatures as $ T_1 \equiv 35 \mu_B (\Delta B_*)/ (2k_B)$ and $ T_2 \equiv 35 \mu_B B_*/(2k_B)$, where $\Delta B_*$ is the separation between $B_*$ and the magnetic field at which the \emph{second} level crossing occurs at $T=0$, namely the second susceptibility peak at $T\to 0$.
As shown by \figref{fig:magnetic_moment_results}(b), the critical temperature $T_1$ corresponds to the temperature above which the step-like behaviour of the magnetic response disappears due to thermal effects.
Finally, at temperatures above $T_2$, the Fermi-Dirac distribution reaches out far above the Fermi level and the discretization of the electronic levels becomes irrelevant, resulting in a magnetic response monotonically dependent on $B_0$.

The various susceptibility regimes described above remain for a wide range of particle sizes $R$, as demonstrated in \cite{SM} where we provide identical figures as \figref{fig:magnetic_moment_results}(b) for $R = 4,5,9$ and $14$ nm. 
However, the critical parameters $B_*$, $T_1$ and $T_2$ decrease with increasing particle size, 
as shown by \figref{fig:BStar_and_Linewidth}(a), where we display $B_*$ and $\Delta B_*$ calculated exactly as a function of particle size (left axis) and the corresponding $T_1$ and $T_2$ (right axis). 
The dependence on the radius $R$ is not smooth since the Fermi level changes abruptly with the radius, but is well fitted by a power dependence $R^{-\alpha}$, with $\alpha = 2.66 \pm 0.62$ and $\alpha=3.94\pm0.99$ for $B_*$ and $\Delta B_*$ respectively.
These values of $\alpha$ reflect a quantum size effect as they can be related to the size dependence of the energy spacing between electronic levels, $E_0\propto R^{-2}$ and the fact that the maximum available values for the quantum numbers $l,m$ increase with increasing particle size.
The decrease of the critical parameters with the particle size hinders, for too large particles, the experimental observation of the exotic magnetic response shown by \figref{fig:magnetic_moment_results}(a).

The magnetic response predicted by our ideal model of isolated, non-interacting electrons is expected to remain in realistic experimental conditions.
Deviations from our model could stem from electron-electron or electron-phonon interactions, surface roughness, or interactions with the particle substrate, among others. 
We describe the effect of all these deviations through the addition of a phenomenological linewidth $\gamma$ to each electronic level \cite{fetter_quantum_2003,breuer_theory_2010}.
Specifically, we weight the original occupation $\bar{n}_{nlms}(T)$ of each single-particle state $\vert nlms\rangle$ with a Lorentzian function, i.e., we modify such occupation to
\begin{equation}
    \bar{n}^\gamma_{nlms}(T) \equiv \int_{-\infty}^\infty \frac{\text{d}E}{\pi} 
    \frac{\gamma \; f_{\rm FD}(E,\mu,T)}{\gamma^2 + (E-E_{nlms})^2},
    \label{eq:linewidth}
\end{equation}
which can be integrated analytically.
The magnetic susceptibility is shown in \figref{fig:BStar_and_Linewidth}(b) for a sphere of $R=3.7$nm as a function of $B_0$ and of the linewidth $\gamma$ normalized to the electronic energy scale for this radius, $E_0= 2.8\text{meV}$.
The peaks in the susceptibility remain for $\gamma\lesssim 10^{-2}E_0$, where individual energy levels are well resolved. The diamagnetic response at $B_0<B_*$ remains visible way beyond this linewidth range.
The presence of a finite linewidth at $T=0$ has, as expected, a very similar qualitative effect as a finite temperature, see \figref{fig:BStar_and_Linewidth}(c) for comparison.

\begin{figure}
	\includegraphics[width=\linewidth]{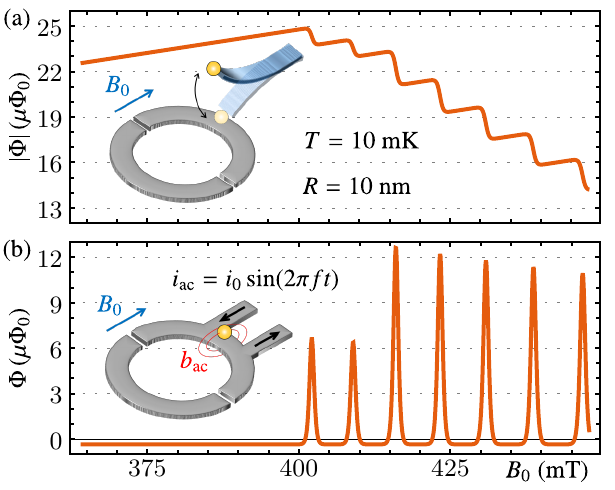}
	\caption{Experimental proposal using a nanoSQUID. 
	(a) The flux induced by a nanoparticle attached to an oscillating cantilever is proportional to the induced magnetic moment $m_z$, Eq.~\eqref{eq:expectation_value_magnetic_moment}. 
	(b) An ac current through the nanoSQUID generates an ac magnetic field at the nanoparticle position, the resulting flux being proportional to the magnetic susceptibility $\chi$, Eq.~\eqref{eq:chi_definition}. In both panels we quantitatively estimate the signal for $R=10$nm, $T=10$mK and typical nanoSQUIDs~\cite{Martinez2017}.}
	\label{fig:Experimental_Flux}
\end{figure}

The quantum size effects shown in this work can be experimentally observed with ultra-sensitive magnetic sensors.
Here we propose two alternative experimental approaches based on YBCO nanoscopic Superconducting QUantum Interference Devices (nanoSQUIDs)~\cite{MartinezRev,GranataRev}, which offer a spin sensitivity of $\sim 10\mu_{\rm B}$/Hz$^{1/2}$ (at $1$ MHz) and can operate at magnetic fields up to $\sim 1$ T~\cite{Schwarz2015,Martinez2017}. In both approaches the signal from the sample is modulated at frequency $f\sim 10-100$kHz in order to exploit the maximum sensitivity of the nanoSQUID, and the dc magnetic field $B_0$ is applied parallel to the nanoSQUID loop and perpendicularly to the plane of the Josephson junctions so that no magnetic flux is coupled to them.
In the first approach (inset \figref{fig:Experimental_Flux}[a]) the nanoparticle is attached to a cantilever and its distance to the nanoSQUID oscillates. This enables lock-in detection of the resulting modulated magnetic flux through the nanoSQUID, whose amplitude is given by $\Phi = \phi_c m_z(B_0)$ with $\phi_c$ the coupling factor~\cite{Martinez2017}. 
As evidenced by \figref{fig:Experimental_Flux}(a), for typical values $R=10$nm, $T=10$mK, and optimum values $\phi_c=10$n$\Phi_0/\mu_B$, with $\Phi_0$ the magnetic flux quantum, the flux experiences steps of  $\sim3\mu\Phi_0$. 
In the second approach (inset \figref{fig:Experimental_Flux}[b]) the particle experiences the oscillating external magnetic field $b_{\rm ac} = b\sin(2\pi f t)$ ($b\ll B_0$) created by an ac current circulating through the nanoSQUID. 
The flux through the nanoSQUID contains an ac component which depends only on the excitation signal (and can be compensated electrically) plus an ac component, $\Phi_{\rm ac}(t) = \Phi\sin(2\pi f t)$, whose amplitude $\Phi = V b \phi_c\mu_0^{-1} \chi(B_0)$ is now proportional to the susceptibility $\chi$  (Eq.~\eqref{eq:chi_definition}). As shown in \figref{fig:Experimental_Flux}(b), for the same typical parameters as above and for $b=5.8$mT~\footnote{According to our simulations, this is the field generated at a height $100$nm above a $200$nm wide$\times80$nm thick nanoSQUID~\cite{Martinez2017} by an ac current of amplitude $i_0=5$mA.}, the flux shows peaks of amplitude $6-12\mu\Phi_0$. Both experimental approaches are feasible as the expected signal exceeds the flux noise floor of YBCO nanoSQUID sensors,  $S_{\Phi}^{1/2} \sim 0.5$ $\mu\Phi_0$/Hz$^{1/2}$ at $f=100$ kHz~\cite{Schwarz2015}.

In conclusion, we have theoretically predicted strong quantum size effects on the ground-state magnetization and magnetic susceptibility of gold nanoparticles of sizes up to tenths of nm, which lie within the measurement capabilities of state-of-the-art magnetic sensing techniques at cryogenic temperatures. 
An interesting outlook of our work consists on studying the thermalization of the electron gas after a deviation from equilibrium. 
This could shed light into the predicted exotic internal energy equilibration in isolated nanoscopic systems~\cite{RubioLopezPRB2018,saavedra_hot-electron_2016}.

We thank J.~J.~Garcia-Ripoll for his valuable suggestions.
C.~G.~B. acknowledges support from the European Union (PWAQUTEC, H2020-MSCA-IF-2017, no.~796725). F. L. and M. J. M. P. acknowledge funding from the Spanish MICINN grant RTI2018-096075-B-C21.

\bibliography{Refs.bib}

\clearpage
\setcounter{figure}{0}
\setcounter{equation}{0}
\counterwithin{figure}{section}
\renewcommand{\theequation}{S.\arabic{equation}}
\renewcommand{\thefigure}{S.\arabic{figure}}
\renewcommand{\bibnumfmt}[1]{[S.#1]}
\renewcommand{\citenumfont}[1]{S.#1}

\onecolumngrid
\begin{center}
	\large{\bf{Supplementary Material for Quantum Size Effects in the Magnetic Susceptibility of a Metallic Nanoparticle}}
\end{center}
\vspace*{10pt}
\twocolumngrid

\section{I. \ Hamiltonian and eigenstates}

\subsection{A. \ Single-particle eigenstates in the absence of magnetic field}

We aim at computing the single-particle eigenstates $\ket{E}$ of
\begin{equation}
    \Hop_0=\sum_{i=1}^N \frac{\ppop_i^2}{2m_e} + U \Theta (|\rrop_i|-R)
\end{equation}
in the limit $U\to\infty$.
In the position representation, these states read $\braket{\rr}{E} \equiv \psi(\rr) \otimes \ket{s}$ where $\ket{s}$ describes the spin state. 
The function $\psi(\rr)$  is given by the solutions of the Helmholtz equation
\hypertarget{eq:radial}{}
\begin{equation}
    -\frac{\hbar^2}{2m_e}\nabla^2 \psi(\rr) = E \psi(\rr).
\end{equation} 
with the boundary condition $\psi(|\rr|\geq R) = 0$.
Eq.~(\hyperlink{eq:radial}{S.2}) can be solved in spherical coordinates by letting $\psi(\rr)=g(r)Y_l^m(\theta,\phi)$ with $Y_l^m(\theta,\phi)$ a spherical Harmonic of order $l$ and $m$.
This leads to the spherical Bessel differential equation for the radial part 
\begin{equation}  \frac{1}{r^2}\frac{\partial}{\partial r}\pare{ r^2 \frac{\partial}{\partial r} g(r)} = 
    \pare{ \frac{l(l+1)}{r^2} - k^2}g(r)
\end{equation}
whose non-singular solutions are given by $j_l(k r)$, with $k^2=\pare{2m_e E}/\hbar^2$ and $j_l(x)$ the spherical Bessel functions of the first kind and order $l$.
The boundary condition imposes that, for a given $l$, $k R$ must be a zero of the spherical Bessel function leading to solutions given by $\psi_{nlm}(\rr)=j_l(u_{ln}r/R)Y_l^m(\theta,\phi)$ where $u_{ln}$ is the $n$-th zero of $j_l(x)$.
The eigenstates $\ket{E}\equiv\ket{nlms}$ are thus characterized by four quantum numbers $n \in \mathds{N}$, $l \in \mathds{N}_0$, $m \in \mathds{Z}$ with $\abs{m} \leq l$, and $s= \pm 1/2$.
They are eigenstates of the orbital and spin angular momentum operators, i.e., [$\hat{l}^2,\hat{l}_z,\hat{s}_z ]\ket{n l m s} = \hbar [ \hbar l(l+1),m,s]\ket{n l m s}$.
The complete normalized eigenstates read in position representation 
\begin{equation}
    \psi_{nlm}(\rr) \otimes \ket{s} = \frac{j_l(u_{ln}\frac{r}{R}) Y_l^m(\theta,\phi)}{\sqrt{N_{nl}}} \otimes \ket{s}
\end{equation}
where $N_{nl}=R^3j_{l+1}(u_{ln})^2/2$ is the normalization constant.
Importantly, the eigenenergies of these states are independent of both $m$ and $s$ and are given by
\begin{equation}
    E_{nl} = \frac{\hbar^2k^2}{2m_e} = \frac{\hbar^2u_{ln}^2}{2m_eR^2}
    \label{eq:Enl}
\end{equation}
which means that, in the absence of magnetic field, all states with the same values of $n$ and $l$ are degenerate.
One refers to each of these manifolds, with degeneracy $2(2l+1)$, as a shell.

\subsection{B. \ Introducing a magnetic field: Minimal coupling}

The Hamiltonian describing the interaction of a charged particle with an external electromagnetic field is obtained, following the minimal coupling scheme \cite{buhmann_macroscopic_2012,sakurai_modern_2014}, by replacing the momentum operator of each particle
$\ppop$ by $\ppop - q\AB(\rrop)$, with $q$ its charge and $\AB(\mathbf{r})$ the magnetic vector potential.
The kinetic energy of a single electron ($q=-e$) in the Coulomb gauge ($\nabla\cdot\AB(\rr)=0$) is thus modified to 
 \begin{equation}
    \frac{\ppop^2}{2m_e} \rightarrow \frac{1}{2m_e} \spare{ \ppop^2 + 2e\,\AB(\rrop)\cdot\ppop + e^2\AB^2(\rrop) }.
\end{equation}
We consider a uniform magnetic field $\BB=B_0\eb_z$, which in the Coulomb gauge corresponds to the vector potential $\AB(\rr)=B_0\frac{\rho}{2}\eb_\phi(\rr)$, with $\rho=r\sin\theta$ the radial distance from the $z-$axis and $\eb_\phi(\rr) = -\sin(\phi)\eb_x+\cos(\phi)\eb_y$ the azimuthal unit vector.
By explicitly writing $\ppop \rightarrow -i\hbar\nabla$ in spherical coordinates one finds
\begin{equation}
    \AB(\rrop)\cdot\ppop \rightarrow B_0\frac{\rho}{2} \pare{ -i \hbar \frac{1}{\rho} \frac{\partial}{\partial\varphi} }
    = \frac{B_0}{2} \hat{l}_z
\end{equation}
which leads to the final expression for the kinetic term
\begin{equation}
    \frac{\ppop^2}{2m_e} \rightarrow\frac{\ppop^2}{2m_e} + \frac{eB_0}{2m_e} \hat{l}_z + \frac{e^2 B_0^2}{8m_e}\rhoop^2.
\end{equation}
Aside from the modification of the kinetic energy, the spin degree of freedom is also affected by the presence of the magnetic field. 
As usual in non-relativistic quantum physics, we must add \textit{ad hoc} the corresponding interaction term in the Hamiltonian, given by
$-\hat{\boldsymbol{\mu}}_e \cdot \BB(\rrop)=eB_0 \hat{s}_z/m_e$
with $\hat{\boldsymbol{\mu}}_e$ the spin magnetic moment of the electron.
Finally, by summing the above expressions over all the electrons, the field-induced correction to the Hamiltonian, $\Hop_B=\Hop - \Hop_0$, is obtained as
\begin{equation}
    \Hop_B = \sum_{i=1}^N \frac{eB_0}{2m_e} \pare{\hat{l}_{z,i} + 2\hat{s}_{z,i}} + \frac{e^2 B_0^2}{8m_e}\rhoop_i^2 .
\end{equation}

\subsection{C. \ Small radius and moderate field approximation}

The single particle Hamiltonian $\Hop$ inside the sphere can be written as
\hypertarget{Seq:single_particle_hamiltonian}{}
\begin{equation}
    \Hop_i = 
    E_0 \pare{\frac{R^2}{\hbar^2} \ppop_i^2 + \eta \frac{\hat{l}_{z,i}+2\hat{s}_{z,i}}{\hbar} + \frac{\eta^2}{4} \frac{\rhoop_i^2}{R^2} }
\end{equation}
where $E_0\equiv\hbar^2/2m_eR^2$ and $\eta \equiv e B_0 R^2 /\hbar$.
In the small particle and moderate field limit ($\eta = e B_0 R^2/\hbar \ll 1$) the last term will contribute much less than the rest and can be treated as a perturbation.
Following standard perturbation theory we rewrite Eq.~(\hyperlink{Seq:single_particle_hamiltonian}{S.10}) as
$\Hop_i = \Hop_i^{(0)} + \epsilon \Vop_i$ with $\epsilon = \eta^2/4$ and $\Vop_i = E_0 \rhoop_i^2/R^2$.
The eigenstates $\ket{nlms^{(0)}}$ of the unperturbed Hamiltonian $\Hop_0$ have unperturbed eigenenergies
\begin{equation}
    E_{nlms}^{(0)} = E_0 \spare{ u_{ln}^2 +\eta (m+2s)}.
    \label{eq:zero_order_eigenenergies}
\end{equation}
Up to first order, the eigenstates $\ket{nlms}$ of $\Hop$ have an energy given by
\hypertarget{Seq:eigenenergies}{}
\begin{equation}
    E_{nlms} = E_{nlms}^{(0)} + \epsilon E_{nlms}^{(1)} + \mathcal{O}(\epsilon^2)
\end{equation}
with $E_{nlms}^{(1)} = E_0 \bra{nlms^{(0)}} \rhoop_i^2/R^2 \ket{nlms^{(0)}}$.
These states up to first order read
\begin{equation}
    \ket{nlms} = \ket{nlms^{(0)}} + \epsilon \ket{nlms^{(1)}} + \mathcal{O}(\epsilon^2)
\end{equation}
with $\ket{nlms^{(1)}}$ given by
\begin{equation}
    \sum_{n'\neq n}\sum_{l'\neq l}
    \frac{E_0}{R^2} \frac{\bra{n'l'ms^{(0)}} \rhoop^2 \ket{nlms^{(0)}}}{E_{nlms}^{(0)} - E_{n'l'ms}^{(0)}} \ket{n'l'ms^{(0)}}.
    \label{Seq:first_order_state}
\end{equation}
where we used $\bra{nlms^{(0)}} \rhoop^2 \ket{n'l'm's'^{(0)}} \propto \delta_{s,s'}\delta_{m,m'}$.


Importantly for this discussion, the magnetic dipole operator $\hat{m}_z$ (see Eq. (4) in the main text) contains the perturbation operator $\rhoop^2/R^2$.
We can make this explicit by expressing the operator as $\hat{m}_z = \hat{m}_z^{(0)} + \epsilon \hat{m}_z^{(1)}$ with 
\begin{align}
    \hat{m}_z^{(0)} = - \frac{\mu_B}{\hbar} 
    \!\pare{ \hat{l}_{z,i} + 2\hat{s}_{z,i}}
    \text{ and }
    \hat{m}_z^{(1)} = - \frac{2\mu_B}{\eta E_0} \Vop_i.
\end{align}

\begin{widetext}
Taking this into account, a consistent perturbative expansion up to first order in $\epsilon$ of $\expect{\hat{m}_z}$ is given by
\begin{align}
    \expect{\hat{m}_z} = \sum_{nlms} \bar{n}\pare{ E_{nlms}^{(0)} + \epsilon E_{nlms}^{(1)}  + \mathcal{O}(\epsilon^2) }
    &\left[ \bra{nlms^{(0)}} \pares{\hat{m}_z^{(0)} + \epsilon \hat{m}_z^{(1)}} \ket{nlms^{(0)}} \right. \\
    &\;\left. + \epsilon \pare{ \bra{nlms^{(1)}} \hat{m}_z^{(0)} \ket{nlms^{(0)}} + \bra{nlms^{(0)}} \hat{m}_z^{(0)} \ket{nlms^{(1)}} }
    + \mathcal{O}(\epsilon^2) \right]
\end{align}
where $\bar{n}(E)$ is the occupation number, given by the Fermi-Dirac distribution.
Here, we don't perform a Taylor expansion of $\bar{n}(E)$ as this function's derivative diverges at $E=0$ as $T\rightarrow 0$.
The expression above can be readily simplified by noting that the term in the second line vanishes.
This is due to the fact that $\ket{nlms^{(0)}}$ is an eigenstate of $\hat{m}_z^{(0)}$ and is orthogonal to $\ket{nlms^{(1)}}$ by construction. Thus, we arrive to the expression used in the main text
\begin{equation}
    \expect{\hat{m}_z} = \sum_{nlms} \bar{n}\pare{ E_{nlms}^{(0)} + \epsilon E_{nlms}^{(1)}  + \mathcal{O}(\epsilon^2) }
    \pare{ \bra{nlms^{(0)}} \pares{\hat{m}_z^{(0)} + \epsilon \hat{m}_z^{(1)}} \ket{nlms^{(0)}} + \mathcal{O}(\epsilon^2) }.
    \label{Seq:correct_expansion}
\end{equation}
\end{widetext}

Ordering $E_{nlms}^{(0)}$ is a challenging task, due to no absolute ordering existing for the spherical Bessel zeros.
Even though there is an analytical expression for the integrals in $E_{nlms}^{(1)}$, the expression still depends on the spherical Bessel functions and their zeros.
Therefore, the inclusion of $\epsilon E_{nlms}^{(1)}$ in Eq.~(\hyperlink{Seq:eigenenergies}{S.12}) further complicates the ordering of the eigenenergies $E_{nlms}$. 
Nevertheless, we have designed an algorithm capable of ordering the eigenenergies efficiently, and subsequently compute the expectation value of the magnetic dipole operator $\expect{\hat{m}_z}$ using the computed ground state.

Our algorithm strongly relies on the partial ordering of the eigenenergies given in Eq.~(\hyperlink{Seq:eigenenergies_relations}{S.21}).
Such partial ordering comes from the intertwining of the spherical Bessel functions zeros and is therefore only guaranteed for the unperturbed eigenenergies $E_{nlms}^{(0)}$.
Nevertheless, although it was designed with the zero-order approximation $E_{nlms} \approx E_{nlms}^{(0)}$ in mind, the algorithm can also be used with the first-order approximation
\begin{equation}
    E_{nlms} = E_{nlms}^{(0)} + \epsilon E_{nlms}^{(1)}.
\end{equation}
as long as one checks that the inequalities in Eq.~(\hyperlink{Seq:eigenenergies_relations}{S.21}) still hold.
This is the case for all the parameter regimes considered in this work. 
Otherwise, as would happen for larger values of $\eta$, one would need to modify the algorithm.

\section{II. \ Computing the induced magnetic dipole} \label{sec:computing_m}

As stated in the main text, in order to compute expectation values of single-particle operators one needs to determine the chemical potential $\mu (T,N,B_0)$ via the following relation
\hypertarget{eq:fermi_energy_relation_SM}{}
\begin{equation}
    N = \sum_{n l m s} \spare{1+\exp \pare{\frac{E_{n l m s}(B_0)-\mu}{k_B T}}}^{-1}=
\end{equation}
where $k_B$ is the Boltzmann constant.
In this section we detail our approach to computing the thermal state of the system, namely the chemical potential $\mu(T,N,B_0)$ and the list of occupied states $\ket{nlms}$ with their respective occupations given by $\bar n_{nlms}(T)$.
We first focus on the case $T=0$ and build our solution in steps.
In Sec.~\hyperlink{subseq:E_F_and_GS}{II.A}, we use the properties of the zeros of the Bessel functions to devise an efficient algorithm to determine all occupied states $\ket{nlms}$ given a value of the Fermi energy $E_F$.
We then derive in Sec.~\hyperlink{subseq:Tilde_E_F}{II.B} an analytical approximation for the Fermi energy $E_F$.
After, in Sec.~\hyperlink{subseq:refine_GS}{II.C}, we show how this approximation can be used as a starting point for a fast numerical computation of the exact ground state at $T=0$, both for $B_0=0$ and for $B_0 \ne 0$.
Finally, in Sec.~\hyperlink{subseq:extend_to_T}{II.D}, we detail the algorithm employed to extend these results to finite temperatures.

Note that the approximation $E_{nlms} = E_{nlms}^{(0)} $ is assumed throughout this section.

\hypertarget{subseq:E_F_and_GS}{}
\subsection{A. \ Ground state and Fermi energy} 

For $T=0$, the Fermi-Dirac distribution reduces to a Heaviside step function $\Theta(E_F(N,B_0)-E_{nlms})$ and therefore solving Eq.~(\hyperlink{eq:fermi_energy_relation_SM}{S.20}) amounts to finding the Fermi energy $E_F(N,B_0)$ and the states $\ket{nlms}$ whose energy fulfills $E_{nlms}\le E_F(N,B_0)$.
In this section we show that if the Fermi energy is known, the list of occupied states can be easily computed.

The algorithm for finding the list of occupied states makes use of the properties of the zeros of the Bessel functions.
Although no absolute ordering relation between these zeros exists, they obey interlacing relations \cite{abramowitz_handbook_2013}, such as $u_{ln}\leq u_{l+1,n}\leq u_{n+1,l}$, which allow to establish the following relations of order between the eigenenergies in Eq.~\eqref{eq:zero_order_eigenenergies},
\hypertarget{Seq:eigenenergies_relations}{}
\begin{subequations}
    \begin{align}
        & E_{n+1,lms} > E_{nlms}  & E_{n,l+1,ms} > E_{nlms} \\
        & E_{nl,m+1,s} > E_{nlms} & E_{nlm,s+1} > E_{nlms}.
    \end{align}
\end{subequations}
In particular, the energy always increases when a single quantum number is increased while leaving the rest unchanged.
With this knowledge, one can devise an algorithm to find the occupied states for a given Fermi energy, that only requires ordering of reduced lists of single-particle energies $E_{nlms}$. 
For each possible value of the spin quantum number $s = \pm1/2$, the algorithm proceeds as follows:
\begin{enumerate}
    \item Find the maximum value of $n$ such that an eigenenergy $E_{nlms}$ with this quantum number and \emph{any} value of $l$,$m$ is below the Fermi energy $E_F$.
    Using the ordering in Eq.~(\hyperlink{Seq:eigenenergies_relations}{S.21}) and $\eta\ll 1$, we determine
    \begin{equation}
        n^s_\text{max} = \max_n \{ n \,\vert\, E_{n,0,0,s} \leq E_F(N) \}.
    \end{equation}
    $n^s_\text{max}$ gives us the maximum value of $n$ for occupied states in the ground state.
    \item For each $n \in [1,n^s_\text{max}]$, find the maximum value of $l$ such that $E_{n lm s}\leq E_F$ for \emph{any} value of $m$. 
    Using the ordering in Eq.~(\hyperlink{Seq:eigenenergies_relations}{S.21}) we determine
    \begin{equation}\label{eq:lmaxs}
        \; l^s_\text{max}(n) = \max_l \{ l \,\vert\, E_{n,l,-l,s} \leq E_F(N) \}.
    \end{equation}
    $l^s_\text{max}(n)$ gives us, for a given $n$, the maximum value of $l$ for occupied states in the ground state.
    \item For each tuple $\{n,l \}$ in the range $1\leq n \leq n^s_\text{max}$ and $0\leq l \leq l^s_\text{max}(n)$, find the maximum value of $m$ such that $E_{n l m s}\leq E_F$. 
    Using the ordering in Eq.~(\hyperlink{Seq:eigenenergies_relations}{S.21}) we determine
    \begin{equation}
        \qquad m^s_\text{max}(n,l) = \max_m \{ m \,\vert\, E_{n,l,m,s} \leq E_F(N) \}.
    \end{equation}
    $m^s_\text{max}(n,l)$ tells us the maximum value of $m$ for occupied states in the shell $n,l$ in the ground state.
\end{enumerate}
In the above way, the ground state is expressed as the set of single-particle eigenstates $\ket{nlms}$ fulfilling $1\leq n \leq n^s_\text{max}$, $0\leq l \leq l^s_\text{max}(n)$ and $-l\leq m \leq m^s_\text{max}(n,l)$ being full and the rest being empty.
Since only few energies need to be ordered, the above algorithm is numerically efficient and it can be applied separately for each particular value of $N$ and $B_0$ provided that the Fermi energy $E_F(N,B_0)$ is known. 
The problem is thus reduced to finding the value of the Fermi energy.

\hypertarget{subseq:Tilde_E_F}{}
\subsection{B. \ Analytical approximation for the Fermi energy at zero field} 

\begin{figure}[t]
    \includegraphics[width=\linewidth]{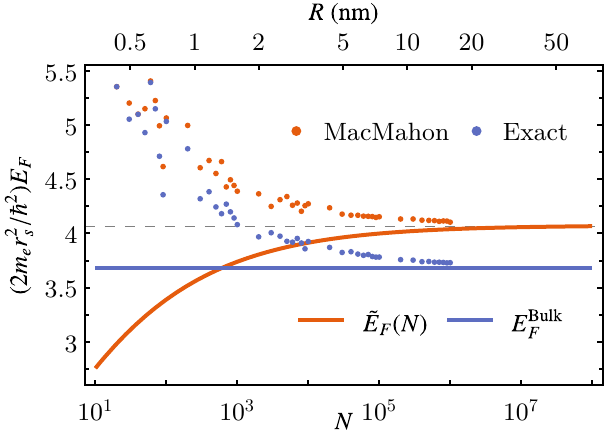}
    \caption{Comparison of different values for the Fermi energy as a function of the number of electrons $N$ or the particle radius $R$ for a gold nanoparticle ($r_s=3 a_0$).
    The orange line corresponds to $\Tilde{E}_F(N)$, the blue line to $E_F^\text{bulk}$ and the blue dots correspond to the exact values of $E_F(N,0)$ found by using the refined approach presented in Sec.~\texorpdfstring{\protect\hyperlink{subseq:refine_GS}{II.C}}.
    The dashed line shows the limiting value of $\Tilde{E}_F(N)$ as $N\rightarrow\infty$.
    The orange dots show how an exact calculation of the ground state as presented in Sec.~\texorpdfstring{\protect\hyperlink{subseq:refine_GS}{II.C}} but using the MacMahon approximation when computing the energies tends to the approximated value $\Tilde{E}_F(N)$.}
    \label{fig:fermi_energies_comparison}
\end{figure}

In the case of zero field and zero temperature, $B_0=T=0$, a useful analytical approximation for the Fermi energy $E_F(N,0)$ can be obtained.
We do this by approximating the sum in Eq.~(\hyperlink{eq:fermi_energy_relation_SM}{S.20}) by the following integral
\begin{equation}
    N \approx \int_{n=1}^\infty \int_{l=0}^\infty \text{d}n \, \text{d}l \, 2(2l+1) \Theta(E_F-E_{nl})
\end{equation}
where the factor $2(2l+1)$, namely the number of states within each shell, is obtained by summing over $m$ and $s$ since the eigenenergies do not depend on these quantum numbers for $B_0=0$.
In order to obtain an analytical expression for this integral, we use the MacMahon approximation \cite{pedersen_stark_2019,abramowitz_handbook_2013} for the spherical Bessel function zeros
\be 
u_{ln} \approx \pare{n+\frac{l}{2}}\pi-\frac{l (l+1)}{\pi  (l+2 n)}.
\ee
Within this approximation, the eigenenergies $E_{nl}$ are monotonically increasing functions of both $n$ and $l$ and the integral of such Heaviside theta function can be written as
\begin{align}
    N \approx 2 \int_{0}^{l_\text{max}} \!\!\text{d}l \int_{1}^{n_\text{max}(l)} \!\!\text{d}n\, (2l+1) \label{eq:approx_N_integral}
\end{align}
where the integration limits are determined via $n_\text{max}(l) \equiv \max_n\{ n | E_{nl}\leq E_F \}$ and $l_\text{max} \equiv \max_l\{ l | E_{1l}\leq E_F \}$ and given by
\be
\begin{split}
    n_\text{max}(l) &= \frac{\sqrt{\varepsilon_F+2 l (l+1)}+\sqrt{\varepsilon_F}-\pi  l}{2 \pi } \\
    l_\text{max} &= \frac{\sqrt{\pi ^2 \varepsilon_F-6 \pi\sqrt{\varepsilon_F}+4 \pi ^2+1}}{\pi ^2-2} \, + \\
    &\quad + \frac{\pi\sqrt{\varepsilon_F}+ 1-2 \pi^2}{\pi^2-2}
\end{split}
\ee
with $\varepsilon_F \equiv E_F/E_0$, where recall that $E_0\equiv\hbar^2/2m_eR^2$.
Performing the integral in Eq.~\eqref{eq:approx_N_integral} yields the following approximate expression for $N$
\begin{equation}
    N \approx \frac{2\pi + 6\pi^2 \sqrt{\varepsilon_F} + 6\pi^3 \varepsilon_F + 8(\pi^2-1)\varepsilon_F^{3/2}}{3 \pi  \left(\pi ^2-2\right)^2}.
\end{equation}
This expression can be inverted exactly to obtain an approximate expression for $E_F(N,0)$, which we denote by $\Tilde{E}_F(N)$.
This approximation to the Fermi energy is shown in \figref{fig:fermi_energies_comparison} as a function of $N$ (solid orange line) and compared to the bulk Fermi energy (solid blue line). 
The dots in \figref{fig:fermi_energies_comparison} correspond to the exact calculation of Fermi energies (see next section for details) computed with the true Bessel zeros (blue dots) and with their approximation according to the MacMahon formula (orange dots). 
For small particle number $N$ the exact Fermi energies are very different both from bulk and from $\Tilde{E}_F(N)$, but in this regime a brute force solution of Eq.~(\hyperlink{eq:fermi_energy_relation_SM}{S.20}) is numerically manageable. 
Conversely, for $N\gtrsim10^6$, the exact solutions converge to the bulk values and the approximation $\Tilde{E}_F(N)$ becomes more accurate. 
Indeed, the above expression shows the correct scaling with the particle density expected for the Jellium model.
Specifically, using $N = (R/r_s)^3$ one finds
\begin{equation}
\begin{split}
    &\lim_{N\rightarrow\infty} \Tilde{E}_F(N) =
    \\
    &=\frac{\hbar^2 N^{2/3}}{2m_e R^2}\spare{ \left(\frac{3\pi(\pi^2-2)^2}{8(\pi^2-1)}\right)^{2/3} 
   \!\! + \mathcal{O}\!\pare{N^{-1/3}} }
    \\
    &= \frac{\hbar^2}{2m_e r_s^2} \spare{ 4.07 + \mathcal{O}\!\pare{N^{-1/3}} }
\end{split}
\end{equation}
whose limiting value for $N\rightarrow\infty$ is in close agreement with the Fermi energy for Jellium bulk \cite{fetter_quantum_2003} given by
\begin{equation}
    E_F^\text{bulk}=\pare{\frac{9\pi}{4}}^{2/3} \frac{\hbar^2}{2m_e r_s^2} \approx3.68\,\frac{\hbar^2}{2m_e r_s^2}.
\end{equation}
The small numerical mismatch  of about $10\%$ is due to the use of the MacMahon approximation for the Bessel zeros. Despite this mismatch, as we will see below, the true power of the analytical approximation resides in providing an excellent initial value for the exact computation of the ground state.

\hypertarget{subseq:refine_GS}{}
\subsection{C. \ Finding the exact ground state at zero temperature}

\begin{figure}
    \includegraphics[width=\linewidth]{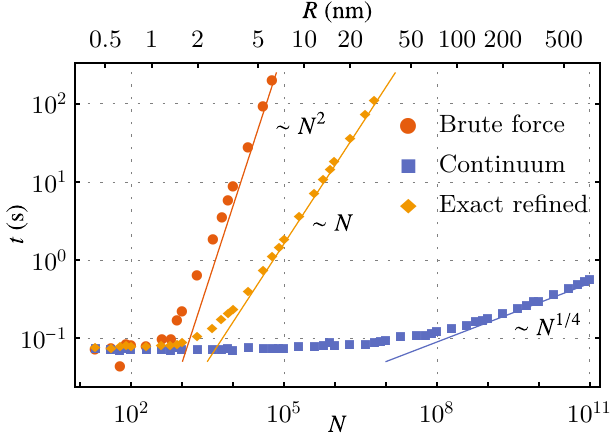}
    \caption{Computation time required, in a tabletop computer, to find the ground state for $B_0=0$ as a function of the number of electrons $N$ or the particle radius $R$ for a gold nanoparticle ($r_s=3 a_0$).
    Red and yellow dots represent, respectively, the brute force solution (see text for details) of Eq.~(\texorpdfstring{\protect\hyperlink{eq:fermi_energy_relation_SM}{S.20}})) and the refined algorithm described in Sec.~\texorpdfstring{\protect\hyperlink{subseq:refine_GS}{II.C}}, both of which allow for an exact determination of the ground state. 
    The blue dots correspond to the approximate solution using the analytical approximation.
    The lines provide a guide to the eye for the approximate scaling with $N$.}
    \label{fig:computation_times}
\end{figure}

In principle, one could use the approximate value $\Tilde{E}_F(N)$ of the Fermi energy to compute the ground state as detailed in Sec.~\hyperlink{subseq:E_F_and_GS}{II.A}, and with it the required expected values.
We will refer to this method as the continuum approximation in the following. 
This, however, is not accurate enough, as one can readily check e.g. by computing the total number of particles through Eq.~(\hyperlink{eq:fermi_energy_relation_SM}{S.20}), 
\begin{equation}
    \sum_{nlms}\Theta(\Tilde{E}_F(N)-E_{nlms}) = \tilde{N} \ne N.
\end{equation}
The significant difference between $\tilde{N}$ and the real number of particles $N$ requires us to refine our computation of the Fermi energy in order to obtain accurate results. 
We thus follow the following algorithm that allows us to compute efficiently the \emph{exact} ground state of the system for any value of $B_0$:
\begin{enumerate}
    \item For each particle number $N$ and magnetic field $B_0$, assume as a first approximation that the Fermi energy is given by $E_F(N,B_0) = \Tilde{E}_F(N,0)(1+\delta)$, with $0.1\leq\delta\leq0.5$ an arbitrary small shift. 
    We remark that using the zero-field analytical expression is a good first approximation for all values of $B_0$ within the moderate field regime we assume in this work.
    \item Compute the ground state following the procedure outlined in Sec.~\hyperlink{subseq:E_F_and_GS}{II.A}. After this we obtain the set of $\tilde{N}\ne N$ lowest energy eigenstates $\vert nlms\rangle$, which corresponds to the exact ground state for a system of $\Tilde{N}$ particles. The addition of the positive shift $\delta$ guarantees that $\Tilde{N}>N$. 
    \item Order the $\tilde{N}$ states in energies and retain the lowest $N$ of the resulting list.
\end{enumerate}

Once the exact ground state is obtained in the above way, the resulting expected values are easily computed, also exactly.
The performance of the above algorithm, in terms of time of computation in a tabletop computer, is shown in \figref{fig:computation_times} (in yellow) as a function of the particle number $N$ and for $B_0=0$ (the timing results depend very weakly on the magnetic field in our regime of interest).
For large $N$, the computation time scales as $N$ and allows to easily calculate the ground state for nanoparticle sizes up to $R \approx 40$ nm. 
This represents a major improvement with respect to a brute force approach where Eq.~(\hyperlink{eq:fermi_energy_relation_SM}{S.20}) is solved by absolute ordering of the eigenenergies Eq.~$\eqref{eq:zero_order_eigenenergies}$.
Such brute force approach, shown in red in \figref{fig:computation_times}, scales as $N^2$ and becomes quickly impractical as $N$ increases.
The advantage provided by our method with respect to the brute force approach stems from the necessity of ordering only a finite set of $\Tilde{N}$ known energies. 
This ordering, however, is still relatively demanding and remains the limiting factor in terms of performance. 
Indeed, note that the computation of the ground state via the continuum approximation, namely performing only steps (1) and (2) in the above enumeration with $\delta=0$, is much more time-efficient as shown in blue in \figref{fig:computation_times}, and allows to obtain results for particles above $R=1\mu$m on a timescale $\sim N^{1/4}$, albeit only approximate. 

We emphasize that our algorithm is, on the one hand, exact, and on the other, not necessarily optimal, although it is numerically manageable for the nanoparticle sizes we focus on. 
For larger spheres, many intermediate algorithms can be devised to compute the ground state and its properties in an approximate fashion, but still largely improving over the analytical expression $\Tilde{E}_F(N)$. 
Investigating this further lies beyond the scope of the present work.

\hypertarget{subseq:extend_to_T}{}
\subsection{D. \ Extension to finite temperature} 

In order to compute results for $T>0$, we find the chemical potential $\mu(T,N,B_0)$ numerically by iteration:
\begin{enumerate}
    \item First, we choose the previously calculated $E_F(N,B_0)$ as an initial guess for the chemical potential $\mu(T,N,B_0)$. This is already a good approximation, since within the considered regime the chemical potential depends weakly on the temperature.
    \item We introduce this guess into the sum in Eq.~(\hyperlink{eq:fermi_energy_relation_SM}{S.20}) and obtain a number of particles $\Tilde{N}^*$, in general different from $N$. 
    The infinite sum is truncated by discarding all the terms smaller than $10^{-16}$.
    \item If $\Tilde{N}^*>N$, the value of $\mu$ is slightly decreased. 
    If $\Tilde{N}^*<N$, the value of $\mu$ is slightly increased.
    \item We repeat steps (2) and (3) iteratively using the Regula Falsi recursive method until the convergence condition $|\Tilde{N}^*-N|<0.5$ is reached.
\end{enumerate}
Although the above method is approximated, it can be used to compute the chemical potential up to arbitrary precision.

\begin{figure*}[ht!]
    \includegraphics[width=\textwidth]{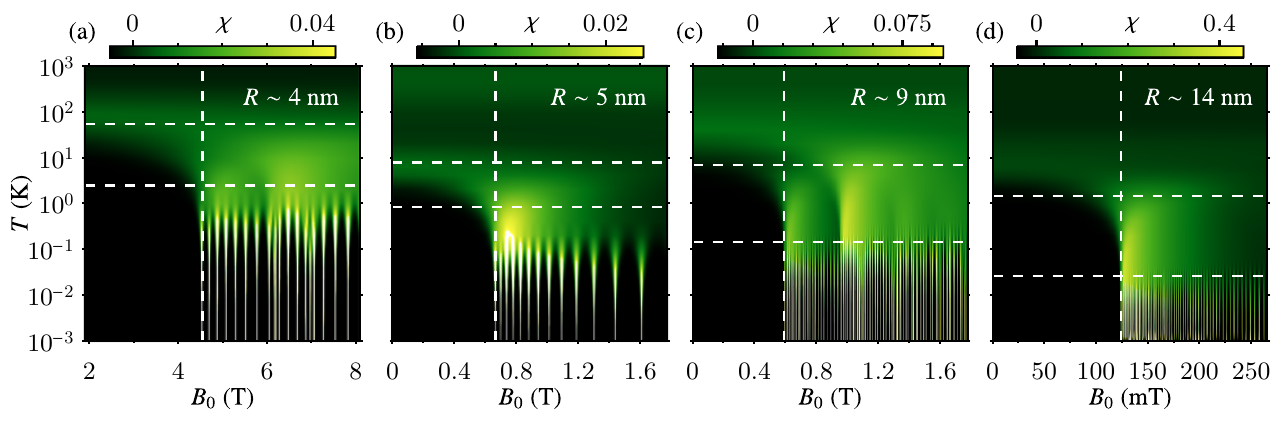}
    \caption{Magnetic susceptibility $\chi$, as defined in the main text, as a function of $B_0$ and $T$ for nanoparticles of different radii.
    The exact number of electrons $N$ in each plot are 12886, 36134, 200126 and 707592, which for gold ($r_s=3a_0$) correspond to $R=$ 3.7, 5.3, 9.3 and 14.2 nm respectively.
    The dashed lines indicate the critical parameters $B_*$, $T_1$, and $T_2$, see text for details.}
    \label{fig:oder_radii_results}
\end{figure*}

\subsection{E. \ Calculation of the magnetic moment}

In order to compute the expected value of $m_z$ we use Eq.~(4) in the main text, which requires to compute the following expected values
\begin{align}
    &\expect{\Lop_z} = \sum_{nlms} \bar{n}_{nlms}(T)\, m, \\
    &\expect{\Sop_z} = \sum_{nlms} \bar{n}_{nlms}(T)\, s, \\
    &\expect{\rhoop^2} = \sum_{nlms} \bar{n}_{nlms}(T)\, \rho^2_{nlm},
\end{align}
where $\rho^2_{nlm}\!\equiv\!\int dV\,\rho^2\,\psi^2_{nlm}(\rr)$ can be integrated analytically using the properties of the spherical Harmonics and the spherical Bessel functions \cite{bloomfield_indefinite_2017}. 
At finite temperature, the sums are carried out over all levels with a relevant occupation $\bar{n}_{nlms}(T)>10^{-16}$.
For $T=0$ the calculation of the above expected values is greatly simplified as the sums in $m$ can be analytically carried out. In the case of the angular momentum and spin operators, 
since most electronic shells are full and since $\sum_{-l}^l m = \sum_{-1/2}^{1/2} s = 0$, one only needs to sum over open shells. 

In order to ease the interpretation of the results, in this paper we have considered only values of $N$ corresponding to fully closed-shell configurations. 
For these values, the magnetic dipole moment at $T=B_0=0$ is exactly $m_z=0$.
Our results, however, hold also for different values of $N$ which would correspond to partially full shells, the difference being the presence of a finite dipole moment $m_z\sim\mu_B$ at $B_0=0$.

\section{III. \ Impact of particle size} \label{sec:More results}

Although in the main text we have focused on a particular radius, our results remain valid for a wide range of particle sizes. 
In \figref{fig:oder_radii_results} we show the magnetic susceptibility as a function of applied field and temperature for four different radii.
Clearly, the magnetic response is strongly dependent on the system size. 
As the size $R$ increases, the quantized jumps in induced magnetic moment, which appear as sharp peaks in the susceptibility, appear at lower applied fields and merge at lower temperatures. 
Despite this strong size dependence, the qualitative behavior of the susceptibility remains the same, namely it shows well defined regions characterized by different magnetic response, and delimited by the three critical parameters introduced in the main text, i.e. $B_*$, $T_1$, and $T_2$.
We can thus quantify the size-dependent magnetic response in terms of the dependence of these three parameters with the particle size.
This dependence is discussed in the main text.

%

\end{document}